# Spectral shaping of fast-gain frequency combs through phases in synthetic dimensions

Diego Piciocchi,[1,2,†,*] Alexander Dikopoltsev,[1,2,†] Ina Heckelmann,[1,2] Mattias Beck,[1,2] Giacomo Scalari,[1,2] and Jérôme Faist[1,2]

[1] Institute for Quantum Electronics, ETH Zürich, Zürich 8093, Switzerland
[2] Quantum Center, ETH Zürich, Zürich 8093, Switzerland
[†] These authors contributed equally to this work.
*dpiciocchi@phys.ethz.ch

**Optical frequency comb devices have unlocked new capabilities in telecommunications, sensing, and metrology. Yet, precise *in situ* control of the comb spectral envelope remains extremely challenging. By introducing mode coupling with non-trivial phases, we demonstrate a spectral shaping technique that enables continuous tuning of a dominant spectral lobe across the full bandwidth of a semiconductor laser frequency comb. We achieve this jointly leveraging the engineered geometry of the synthetic lattice formed by the cavity modes of the laser and the coherent dynamics enabled by its fast-gain recovery. We use dual-tone modulation of the cavity at its repetition rate and twice this frequency with a controlled relative phase to couple the comb modes into a triangular lattice. The relative phase between the two tones defines a lattice phase that breaks time-reversal symmetry and steers the lattice dynamics through the fast gain. With this approach, we experimentally control the spectral envelope of the comb such that a targeted region contains more than twice the intensity expected from a uniform distribution, demonstrating tunable spectral selectivity. This capability, achieved directly at the light generation stage in a fast-gain device, opens routes for efficient programmable waveform engineering with potential applications in ranging, data transmission, and sensing.**

## 1. INTRODUCTION

Over the past decades, optical frequency combs (OFCs) have driven innovation across science and engineering [1], enabling breakthroughs in spectroscopy [2], metrology [3], and microwave synthesis [4]. The development of compact platforms [5], such as micro-combs and mode-locked lasers, has significantly reduced the size and power consumption of OFC sources, enabling portable applications and foundry compatibility [6]. These advances unlock dense data transmission [7], parallelized photonic neural networks [8] and on-chip sensing [9], all within a miniaturized form factor.

Controlling an OFC's spectral intensity distribution is valuable across multiple fields. In spectroscopy, spectral reconfiguration of the source allows to tune the response function and removes complexity from the detection system, enabling single-pixel spectroscopy [10]. For applications using comb lines as orthogonal sources, such as dense optical communication [11] or ranging [12], frequency-domain synthesis enables power management and channel navigation. In quantum information, spectral control improves atomic transition control [13], superconducting qubit driving [14], and frequency-bin encoding [15]. Existing methods to shape the output spectrum of a device rely on dispersion engineering [16,17] or on-chip filters [18,19]. However, portable systems would benefit from *in situ* shaping, which reduces complexity and allows for spectral reconfiguration within the light source during operation.

Due to their perfect equal spacing, the modes of an OFC can be viewed as sites in a synthetic frequency lattice [20,21], enabling the exploration of phenomena such as Bloch oscillations [22,23], topological insulation [24], gauge potentials [25], non-Abelian fields [26], and photonic analogs of the Quantum Hall effect [27]. Lattices in synthetic dimensions offer key advantages over real space systems, such as tunable long-range interactions [28] or increased dimensionality [29–31], and hold the key to precise and robust spectral control. Recently, synthetic lattices have found applications in efficient quantum simulations [32], optical isolation [20], and theoretical proposals for mode-locking [33]. Most works on synthetic dimensions are conducted in linear optical systems, which lack the intrinsic nonlinear mechanisms required for the stabilization of cavity-based frequency comb states. Hence, approaches that fully exploit this engineering flexibility to shape OFC spectra in active devices remain scarce [22,34].

We propose and demonstrate an efficient *in situ* method for shaping OFC spectra by controlling coupling phases in the synthetic frequency lattices of modulated fast-gain ring lasers. This capability emerges from the dual-tone modulation that generates a non-trivial phase in the coupling between cavity modes [Fig. 1(A)] and is stabilized by the coherent liquid-like dynamics of the intracavity field. Modulating the laser cavity at its resonances $f_{rep}$ and $2f_{rep}$, we induce nearest- and next-nearest-neighbor coupling along the synthetic frequency lattice of cavity modes, forming a triangular ladder geometry [Fig. 1(B)]. The relative phase between these couplings defines a lattice phase that breaks time-reversal symmetry, modifying the band structure [Fig. 1(C)] and directing energy transfer between lattice sites. The fast-gain recovery in our system suppresses intensity fluctuations on the fastest timescale of the system, ensuring quasi-constant intensity and granting coherent flow dynamics to the intracavity field [23] [Fig. 1(D)]. This in turn also maximizes the overlap between the temporally extended field and the modulation, making the spectral dynamics highly sensitive to the engineered lattice and its coupling phase. Ultimately, the liquid-like behavior enables the response to external control, allowing dynamic spectral shaping directly at the light generation stage [Figs. 1(E), 1(F)].

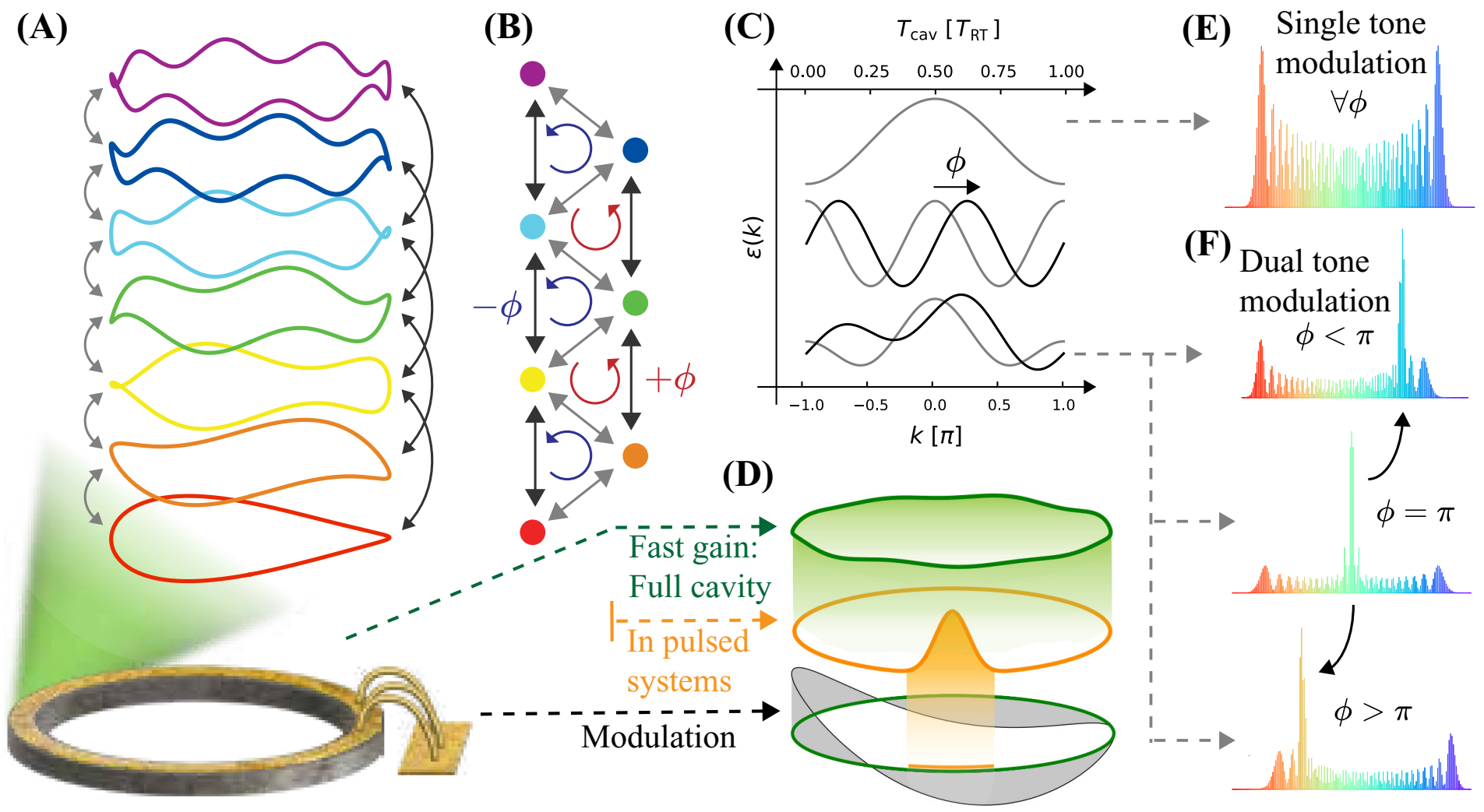

**Fig. 1.** Comb shaping with synthetic lattice phases in fast-gain lasers. (A) The synthetic lattice is realized using the cavity modes of a ring-shaped laser. (B) Nearest- and next-nearest-neighbor couplings are introduced via modulations at $f_{rep}$ and $2f_{rep}$, with their relative phase imparting a path-dependent phase in the lattice. (C) The modulation profile determines the lattice's band structure. Breaking time-reversal symmetry requires two modulation tones with a relative phase. (D) Our fast-gain system, with nearly constant intensity, probes the modulation over the entire roundtrip—unlike pulsed systems, which sample only a localized portion. (E) With a single modulation tone, the lattice phase has no impact on the spectrum. (F) With dual-tone modulation, the system's steady-state behavior is governed by the lattice phase, enabling spectral shaping.

Our demonstration, based on the continuous wave operation of fast-gain devices, clarifies why most pulsed OFC sources have yet to fully leverage the potential of synthetic dimensions for spectral control. Although electro-optic combs [35] achieve flexible spectral shaping by coupling OFC modes through amplitude and phase modulation, they do not operate in a resonant regime. Therefore, the nonlinear process requires long modulation paths and significant power [36], on the order of a few watts [37]. Resonant electro-optic schemes are more power-efficient but suffer from limited controllability due to the absence of nonlinear stabilization mechanisms [38], while pulse-generating nonlinearities further reduce electro-optical conversion efficiency [39]. Our technique enables on-the-fly spectral tuning directly within portable OFC sources, leveraging fast gain [40–44] and reducing the modulation power through cavity resonant processes, opening new possibilities in ranging, spectroscopy, and communications.

## 2. PHASES IN SYNTHETIC LATTICES

We consider a laser with a circular cavity [45,46], whose modes define a one-dimensional lattice along the synthetic frequency dimension [Fig. 2(A)]. The modes are coupled by radio-frequency modulation of the injected current at harmonics of the cavity resonance frequency, $N \cdot f_{rep}$. The latter is translated into phase modulation through the linewidth enhancement factor [47], enabling coupling between $N^{\text{th}}$-neighbors along the lattice. Here, we introduce a coupling phase in the lattice by combining radio-frequency (RF) modulations at $f_{rep}$ and $2f_{rep}$, and controlling their amplitudes $A_1$, $A_2$ and their relative phase $\phi$.

Using this modulation scheme, we induce nearest-neighbor (NN) and next-nearest-neighbor (NNN) couplings [Fig. 2(A)] in the synthetic lattice [48]. The onsite potential is quadratic and complex and is determined by the dispersion, $\beta$, gain curvature, $g_c$, and cavity wavevector $K$. Consequently, the linear part of this system can be described using the following equations:

$$i\frac{\mathrm{d}B_m}{\mathrm{d}\tau} = (D - iG)m^2 B_m + C_{\mathrm{NN}} \cdot (B_{m+1} + B_{m-1})$$
$$+ C_{\mathrm{NNN}} \cdot (e^{i\phi} B_{m+2} + e^{-i\phi} B_{m-2}), \qquad (1)$$

where $B_m$ is the complex-valued occupation of the $m$-th synthetic lattice site (i.e., amplitude of the $m$-th mode), $D = \beta K^2/2$ and $G = g_c K^2/2$. $C_{\mathrm{NN}} = A_1/2$ and $C_{\mathrm{NNN}} = A_2/2$ are the amplitudes of the NN and NNN coupling coefficients, respectively (Section 1 of Supplement 1). The relative phase $\phi$ between the two modulation components is translated to the phase factor $e^{i\phi}$ for the coupling term $C_{\mathrm{NNN}}$ and cannot be trivially removed. Equation (1) can be written in a more compact form using a Hamiltonian operator on the right side: $i\frac{\mathrm{d}B}{\mathrm{d}\tau} = HB$ (Section 1 of Supplement 1). This Hamiltonian describes the structure of a lattice with a triangular ladder geometry, which is shown in Fig. 2(A), with even and odd modes appearing on different sides. The phase $\phi$ determines the total non-trivial phase accumulated around each lattice plaquette. This can be visualized by following a closed hopping loop—comprising two nearest-neighbor hops and a next-nearest-neighbor one, whose orientation (clockwise or counterclockwise) yields an accumulated phase of $+\phi$ or $-\phi$, respectively [Fig. 2(A)].

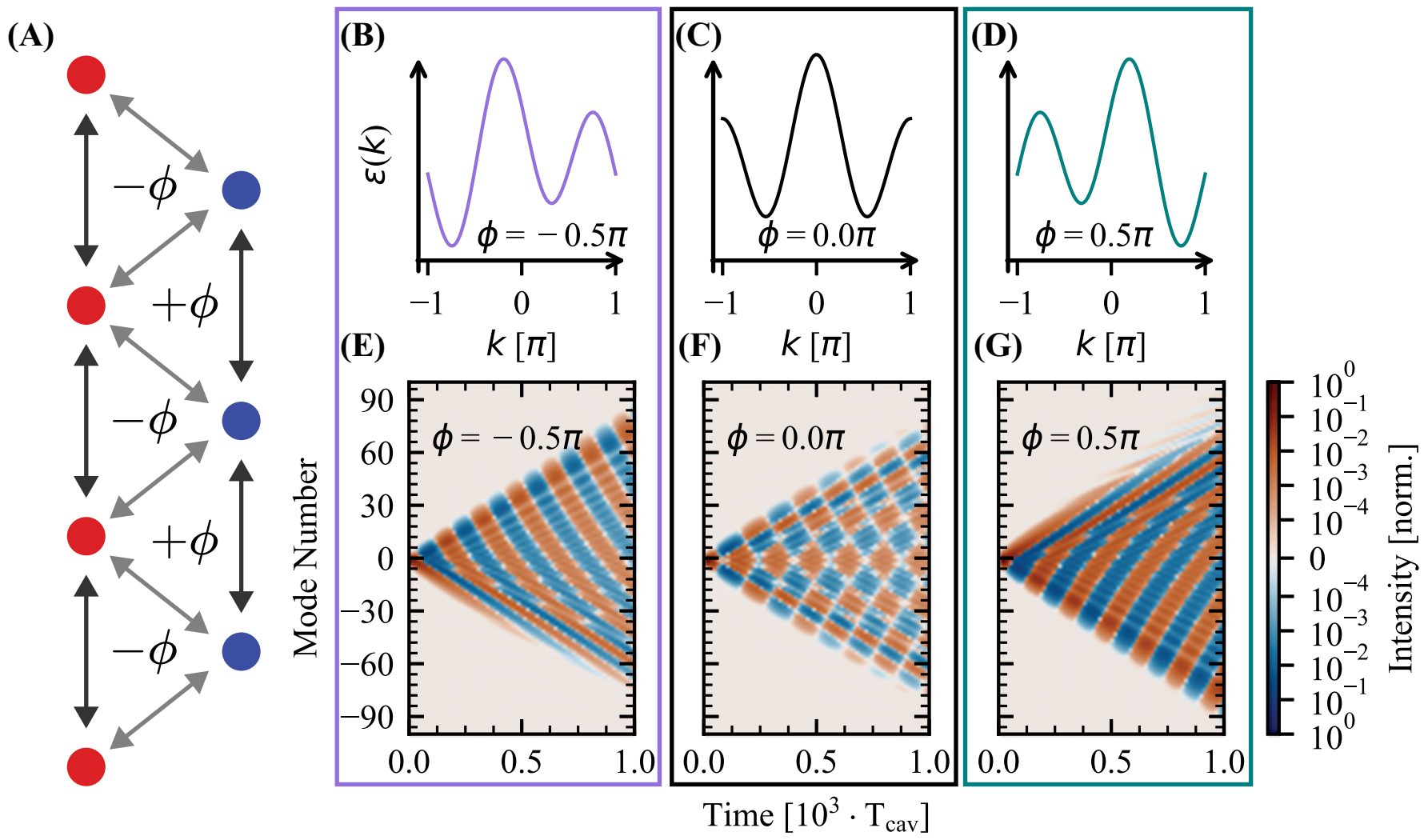

**Fig. 2.** Linear 1D lattice dynamics with a nontrivial phase. (A) The triangular lattice geometry with a nontrivial phase, shown with even lattice sites on the left (red) and odd ones on the right (blue). (B–D) The band structure of the synthetic lattice, dictated by the modulation in the cavity's real space (the synthetic lattice's reciprocal space). (E,F) Time evolution of a single-site excitation in triangular ladder lattices with different lattice phases ($\phi = 0, \pm\pi/2$), obtained by solving the coupled wave equation [Eq. (1)] with $D = G = 0$. With this configuration, which includes only nearest and next-nearest neighbor coupling but lacks a confining potential, the wavefunction expands infinitely and does not lock into a steady state (see Section 2 of Supplement 1). When $\phi \neq 0, \pi$, time-reversal symmetry is broken, causing directional energy transfer between odd (blue) and even (red) sublattices.

The lattice phase effectively breaks the time-reversal symmetry of the system [48]. Specifically, the modulation at $f_{\text{rep}}$ and $2f_{\text{rep}}$, which is defined in the real space of the system, causes coupling along the synthetic frequency dimension, i.e., in the system's reciprocal space. The band structure of this synthetic lattice is defined in its reciprocal space, which in this case is again the real space of the cavity. Interestingly, this implies that the modulation directly corresponds to the band structure, given by

$$\varepsilon(k) = 2C_{\text{NN}} \cos(k) + 2C_{\text{NNN}} \cos(2k + \phi). \quad (2)$$

As explained above, the quasimomentum coordinate $k$ used here corresponds to the corotating frame coordinate $Kz$. In the laboratory frame, this becomes the cavity time, used in the top axis of Fig. 1(C). Breaking of the time-reversal symmetry is reflected in the fact that as $\phi \neq 0, \pi$ the band structure $\varepsilon(k)$ is no longer symmetric around $k = 0$ [Figs. 2(B)–2(D)] and impacts the population dynamics of the lattice. Figures 2(E)–2(G) illustrate the isolated effect of the phase excluding both the quadratic potential (i.e., $D = G = 0$) and nonlinear gain dynamics, obtained solving Eq. (1). Starting from a single lattice site, the wavefunction expands, in principle, to infinity, without stabilizing on a single steady state that can be used as a frequency comb. Panels E, F, and G show the dynamics of the lattice population for a trivial phase $\phi = 0$ [Fig. 2(F)] and nontrivial phases $\phi = \pm 0.5\pi$ [Figs. 2(E), 2(G)]. In the regime of broken time-reversal symmetry, we observe a periodic intensity between the even (red) and odd (blue) sub-lattices, with the direction of transfer dependent on the sign of the lattice phase $\phi$. Instead, a symmetric band structure with a trivial phase $\phi = 0$ [Fig. 2(C)] shows symmetric dynamics between the even and odd sub-lattices [Fig. 2(F)]. Figure 2 illustrates this behavior only for three values of the lattice phase $\phi$: the evolution of this exchange as well as the band structure is shown for all values of $\phi \in [0, 2\pi]$ in Section 2 of Supplement 1, presenting a self similar expanding distribution.

## 3. ENABLING SPECTRAL CONTROL WITH FAST GAIN

The coupling phase dynamics are imprinted on the steady-state of our mode-locked laser source. The ring-shaped semiconductor laser is based on a quantum cascade laser (QCL) [45] active medium with an ultrafast-gain recovery time, <1 ps, much shorter than the cavity round-trip time, which is on the order of 10s of picoseconds [44]. Such gain adds a non-Hermitian nonlinear term to the field dynamics. This is included in the equations for the evolution of the mode amplitudes $B$ as

$$i\frac{\mathrm{d}B}{\mathrm{d}\tau} = H \cdot B + F_{NL}(I(\tau)), \quad (3)$$

where $I(\tau)$ is the time-dependent intensity of the electromagnetic field. The solutions of this equation, including both a confining quadratic potential $D \neq 0$ and a fast-gain nonlinearity, provide the steady states for the lattice occupation. These correspond directly to the intensity in each mode at frequency $\nu_m$, i.e., the spectrum of the device $S(\nu_m) = |B_m|^2$.

The role of the fast-gain nonlinear term $F_{NL}$ (Section 3 of Supplement 1 for details) is to induce non-Hermitian long-range coupling between the modes that forces quasi-constant intensity in the cavity. We note that our system naturally fulfils the condition of starting at a single site, since the gain causes spontaneous symmetry breaking of the lasing direction, while the low surface roughness enables single mode lasing just above threshold [43]. When phase modulation is applied, the coupling transfers population to other sites of the synthetic lattice, while the fast-gain constrains the amplitudes and phases to keep the total intensity quasi-constant in time [43]. This suppression of intensity fluctuations on the fastest timescale of the system is what endows our intracavity field with the coherent liquid-like dynamics (Section 4 of Supplement 1), and eventually enables the coupling phase-driven spectral shaping. The resulting intracavity intensity, being approximately constant with time, extends over the whole cavity cycle and overlaps maximally with the modulation signal, as shown in Fig. 3(A). This makes the intracavity field sensitive to the functional shape of the modulation across the full period. This is in contrast to a pulsed intracavity field that will only probe a local part of the modulation shape, here approximately parabolic, being insensitive to its symmetry properties for the rest of the period [22,39]. Therefore, the addition of the fast-gain term is crucial for enabling sensitivity to the underlying synthetic lattice. This fast-gain-enabled sensitivity to an external electrical signal is what differentiates our work from previous studies on self-starting frequency combs from fast-gain lasers [47,49]. In self-starting combs, comb formation is driven by intrinsic processes like spatial-hole burning, and not by an external signal. Consequently, the shape of the comb is determined by the device's inherent characteristics, such as dispersion or gain curvature, and cannot be externally controlled.

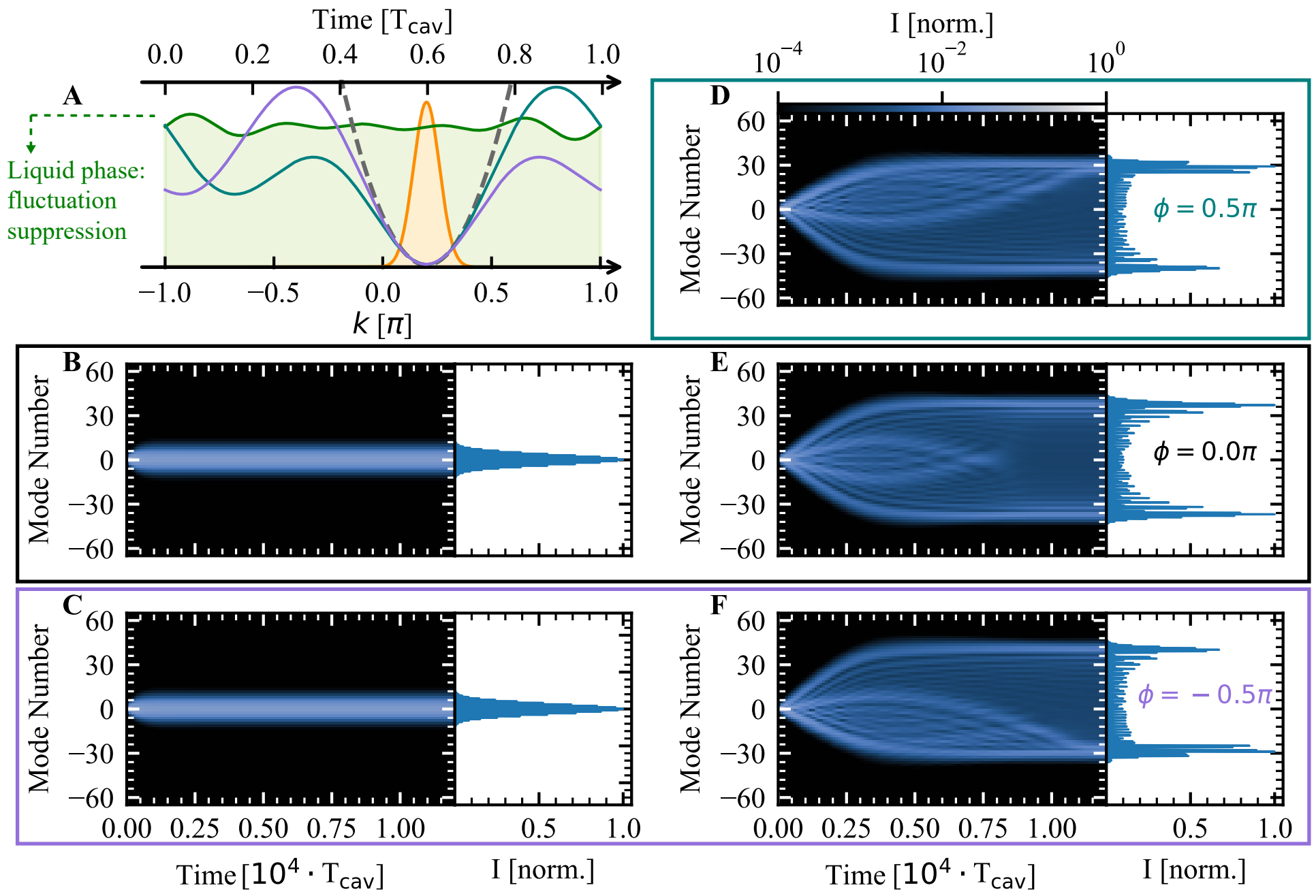

**Fig. 3.** Interplay between lattice phases and liquid light. (A) Comparison of pulsed and extended signals under modulation. Fast gain ensures a quasi-continuous intracavity intensity that overlaps with the modulation profile throughout the whole round-trip. In contrast, systems with pulsed intensities only sample the local shape of the modulation, which here is a parabolic minimum. This explains why pulsed signal are only weakly manipulated by phase modulation. (B), (C) Time evolution of the spectral state obtained solving Eq. (3) with slow gain and $D$, $G \neq 0$, supporting pulsed lasing. We observe that for two different modulation signals, the resulting steady state is insensitive to the phase $\phi$. (D–F) Time evolution of the spectral state obtained solving Eq. (3) with fast gain and $D$, $G \neq 0$, showing a significant response to the change in $\phi$. By probing the full modulation, both the dynamics and the steady state show the chiral lattice dynamics, making the final spectrum sensitive to $\phi$. We note that in the absence of a confining potential, the system would not stabilize on a single state and could not be used as a frequency comb.

To show the impact of this sensitivity to the modulation, we compare dynamics in the fast-gain medium to media in which the gain recovery is slow, where the losses from saturation act on the averaged signal, i.e., $g(\tau) = g_0/(1 + \langle I(z, \tau)\rangle_z / I_{sat})$. Figures 3(B), 3(C) show the time evolution of the spectrum of slow-gain ring lasers using modulations with the same amplitudes for the $f_{rep}$ and $2f_{rep}$ components and two different values of $\phi = 0, \pi/2$. Even though the two lattice phases produce modulation profiles that differ with respect to their time-reversal symmetry properties, the profile around the minimum is almost unaffected. In both cases, the evolution starts from a single mode, corresponding to the occupation of a single lattice site. At time $t = 0$, the modulation is switched on and the modes begin to proliferate. With slow gain, the stabilization is governed by dissipation, induced by gain curvature. This process is also known as active mode locking, and it typically leads to a spectrally narrow Gaussian state [50] [Figs. 3(B), 3(C)]. In contrast, the continuous intensity enforced by the fast gain makes the system sensitive to the full modulation profile. As a consequence, the directional transfer of energy in the underlying lattice [Figs. 2(E)–2(G)] is imprinted onto the nonlinear dynamics, as shown in Figs. 3(D)–3(F) for three values of the phase, $\phi = 0, \pm\pi/2$. The nontrivial phase influences both the initial dynamics and the steady state, as is particularly evident when symmetry is broken ($\phi = \pm\pi/2$). Crucially, the steady state is a result of the combined action of the fast gainwhich enables the liquid-like dynamics (Section 4 of Supplement 1) and the quadratic confining potential. Without this potential, the bandwidth would continuously expand, preventing stabilization onto a steady state and spectral shaping as a consequence. When both of these effects are present, the steady state of the system is exceptionally sensitive to engineered synthetic lattice geometry, which makes *in situ* spectral control possible. Indeed, in the complete nonlinear system including gain dynamics and dispersion, the modulation waveform is encoded in the instantaneous emitted frequency, which ultimately shapes the spectral envelope (see the analytical solution discussed in Section 5 of Supplement 1).

## 4. DEMONSTRATION OF SPECTRAL SHAPING

The ring-shaped QCL has a cavity length of $L_c \approx 7.1$ mm, corresponding to a resonance frequency of $f_{\text{rep}} = 12.534$ GHz. When operated without any modulation, the device is in the single mode regime, emitting at $\nu_0 = 1341\ \text{cm}^{-1}$ ($\approx 7.46\ \mu$m). We demonstrate control over the spectrum by manipulating the synthetic lattice with two synchronized radio frequency current modulation signals at $f_{\text{rep}}$ and $2 f_{\text{rep}} = 25.068$ GHz, injected into the semiconductor fast-gain ring laser device, and translated to phase modulation through the gain. For each value of the phase between the two modulation frequencies $\phi$, we allow the laser to reach the steady state spectrum and measure it with a Fourier transform infrared (FTIR) spectrometer (details in Section 6 of Supplement 1).

Figures 4(A), 4(B) show the effect of the amplitude of the two modulation components on the steady state spectrum for a fixed value of $\phi = \pi$ in both experimental measurements and numerical simulations (see Section 7 of Supplement 1 for a continuous scan of the values of $\phi$). An increase in amplitude $A_2$ (or $C_{\text{NNN}}$) leads to the progressive appearance of a peak in the central region of the spectrum, demonstrating that tuning the amplitudes of the coupling enables the observation of localized features in the spectrum. In particular, the bandwidth achieved ($\approx 15\ \text{cm}^{-1}$) is constrained by the power limitations of the source at $2 f_{\text{rep}}$ and does not represent an intrinsic limitation of the driving scheme (see Section 8 of Supplement 1 for more details). By increasing the radio-frequency modulation power through the RF optimization, dual-tone modulated states offer the potential of reaching the bandwidth of Quantum Walk Comb states, which is on par with state-of-the art mid-IR [51] and THz QCL combs [52], as well as quantum dot [42] and quantum well lasers [53].

Next, in Figs. 4(C), 4(D), we show that this feature can be manipulated by changing the lattice phase across the $[0, 2\pi]$ range, demonstrated for two different values of $A_{1,2}$. The resulting steady state spectral shapes contain peaks that scan the entire span of the available bandwidth, allowing for controllable intensity transfer across the entire spectral range. We also note that, as the ratio between the NN and NNN coupling decreases, the intensity transfer in the spectrum becomes progressively smoother, illustrating the combined effect of shaping using the two degrees of freedom of relative amplitudes and lattice phase. The good quantitative agreement with the numerical simulations [Fig. 4(E)], based on the model in Eq. (3), confirms that the lattice phase breaks time-reversal symmetry and directly affects the output spectrum of the device. Because the modulation is transferred to the steady state via the instantaneous frequency, this imprinting requires a well-defined optical phase and thus a coherent interplay between the comb modes (Section 7-A of Supplement 1). We further note that the model—which includes only modulation, a confining potential, the lattice phase, and fast gain—captures the fundamental physics underlying the observed spectral shaping (discussed in Section 3 of Supplement 1).

We have shown that the lattice phase can direct the comb's power to a chosen frequency range. Next, we quantitatively evaluate the spectral shaping. To do so, we introduce a metric for spectral shaping, which we call the spectral steering capacity ($\eta$, details in Section 9 of Supplement 1). This quantity represents the maximum fraction of total power that can be moved across the whole spectrum by adjusting a control parameter (here, it is $\phi$), scaled by the number of frequency channels, $N$. The spectral steering capacity, $\eta_N$, reflects the increase in the ratio between the signal in a specific frequency range and the background in the rest of the spectrum. To achieve control, this value must be $> 0$. Importantly, this quantity is not equivalent to an SNR, as the background in this case is a systematic signal that cannot be averaged out. With this definition, any non tunable spectrum has a steering capacity $\eta_N \leq 0$, where the value zero corresponds to a uniformly distributed spectrum. A value $\eta_N > 0$ indicates controllable spectra. Using the lattice phase $\phi$ as the tuning parameter, our spectral shaping scheme redistributes up to 30% of the total comb power across eight frequency channels ($\eta_8 \approx 1.26$) in simulations, and more than 20% ($\eta_8 \approx 1.15$) in experiments. This steering capacity is substantially higher than the one achievable by detuning a single RF injection tone from resonance, which is limited to $\eta_8 \approx 0.45$ (Section 9 of Supplement 1).

Finally, to evaluate the stability of the shaped frequency comb states, we measured their amplitude-noise in the low frequency range, from 1 kHz to 20 MHz. We find that this modulation scheme, enabling spectral shaping, does not degrade the laser performance in this respect (Section 10 of Supplement 1). To estimate the noise properties at higher frequencies, we performed simulations concerning the noise response of the intracavity field (discussed in Section 4 of Supplement 1). These illustrate that, even with the dual-tone modulation scheme, noise is substantially reduced compared to systems with slow-gain media.

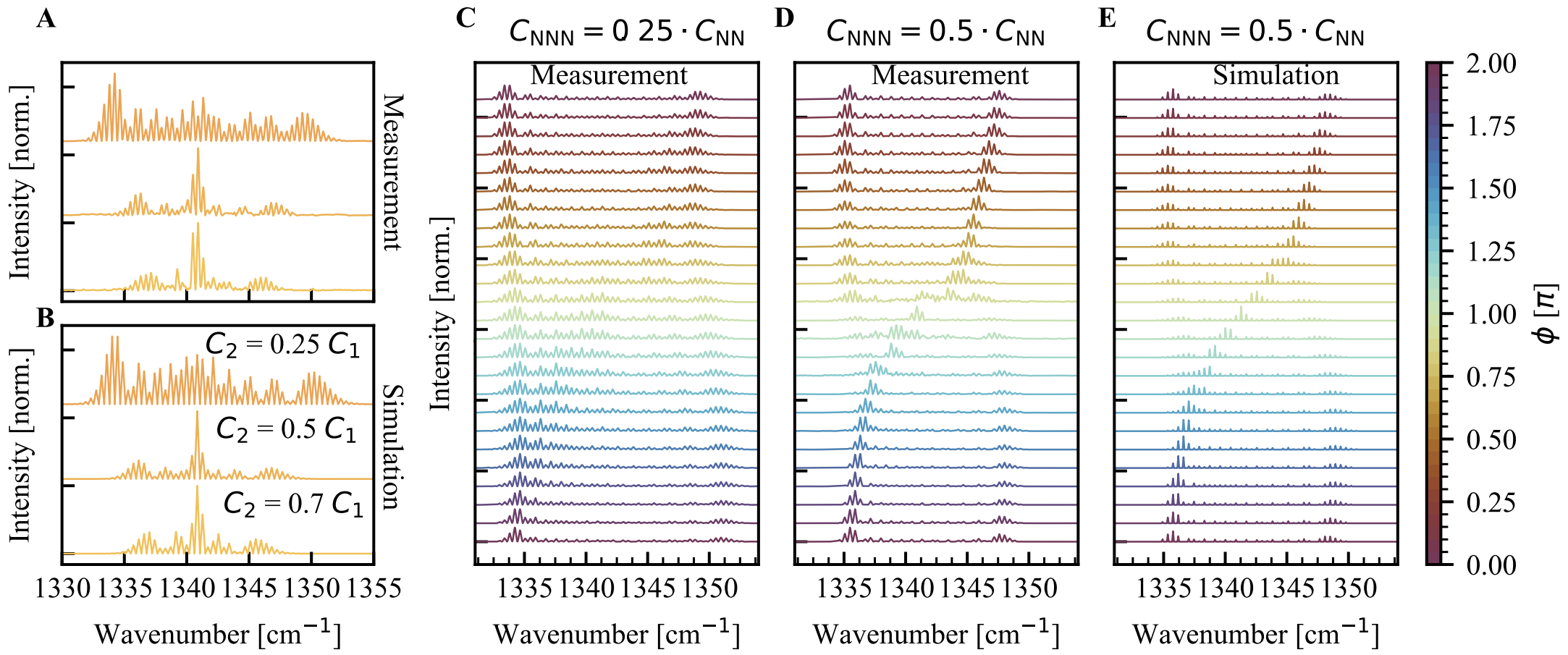

**Fig. 4.** Spectral shaping with a lattice phase. (A,B) Effect of changing the amplitudes of the signals at $f_{\text{rep}}$ and $2 f_{\text{rep}}$ for a fixed lattice phase $\phi = \pi$. As the amplitude of the $2 f_{\text{rep}}$ component increases, a peak in the central region of the spectrum appears. (C, D) Effect of the phase, $\phi$, between the modulation signals at $f_{\text{rep}}$ and $2 f_{\text{rep}}$ on the steady-state optical spectrum of the fast-gain laser. As $\phi$ is increasing, the peak from (A), (B) is changing its central position to higher frequencies in the spectrum, until it winds back. The injection components have fixed values of their relative amplitudes, with $C_{\text{NNN}} = 0.25 C_{\text{NN}}$ and $C_{\text{NNN}} = 0.5 C_{\text{NN}}$, respectively. (E) Simulation of the steady state of the same amplitude modulation values in (D).

## 5. CONCLUSION

We have demonstrated a spectral shaping scheme based on the engineering of a triangular synthetic lattice with a lattice phase, producing configurations that break time-reversal symmetry. This is achieved by modulating the current of a mid-IR QCL at the cavity's free spectral range and its second harmonic. The fast-gain recovery of the semiconductor device forces a quasi-constant field intensity over the entire optical cycle, i.e., liquid-like light [23,54]. This property allows the nonlinear system to access new steady states that inherit the properties of the underlying synthetic lattice, enabling control of the spectral envelope by tuning the relative amplitude and phase of the two modulation components. We have quantitatively shown that this approach provides direct *in situ* control of the spectrum that surpasses regular comb formation techniques. We also found that the noise properties of the comb do not degrade with multiple-frequency injection, which holds the potential for highly stable controllable frequency comb devices.

Already with the dual-tone modulation scheme demonstrated here, tunable frequency combs can reduce the complexity of spectroscopic systems by shifting spectral manipulations from interferometric detection to direct control at the light source [55]. Furthermore, this technique could be applied to pulse compression by dispersion compensation [56], transferring the flexibility of spectral tuning to pulse shaping.

A natural extension of this work is the implementation of modulation schemes at higher harmonics [31,57] to introduce $N$th-neighbor couplings in the synthetic lattice. This would translate fine-grained RF domain signal engineering into the optical domain, paving the way for a new generation of optical synthesis techniques. Moreover, Quantum Walk Comb operation is supported by any gain medium with instantaneous incoherent pumping. Notably, recent work has demonstrated Quantum Walk Comb operation in an interband device operating near the C band with a low repetition rate cavity, ensuring efficient injection up to $13 f_{rep}$. Such platforms offer particularly promising opportunities for flexible spectral shaping of optical frequency combs in communication-relevant wavelength regimes.

Together, these perspectives underscore the potential of multi-frequency modulated fast-gain laser as tunable and customizable light sources.

**Funding.** Staatssekretariat für Bildung, Forschung und Innovation (22.00182, MIRAQLS, EU grant agreement, 101070700); Schweizerischer Nationalfonds zur Förderung der Wissenschaftlichen Forschung (212735); Innosuisse-Schweizerische Agentur für Innovationsförderung (52899.1 IP-ENG, 2155008433, "High yield QCL Combs"); Eidgenössische Technische Hochschule Zürich (22-1 FEL-46, ETH Fellowship program).

**Acknowledgment.** D.P. and A.D. performed the characterizations, developed the models, performed the numerical simulations, and wrote the original draft. A.D. and I.H. conceptualized the idea. D.P. wrote the Supplement 1. I.H. processed the devices. M.B. grew the QCL wafer and consulted in the fabrication. A.D., G.S., and J.F. acquired the funding, administrated, and supervised the project. All authors contributed to the interpretation of the results and the review and editing of the draft.

**Disclosures.** The authors declare no conflicts of interest.

**Data availability.** Data underlying the results presented in this paper are available in Ref. [58].

**Supplemental document.** See Supplement 1 for supporting content.

## REFERENCES

1. S. A. Diddams, K. Vahala, and T. Udem, "Optical frequency combs: coherently uniting the electromagnetic spectrum," Science **369**, eaay3676 (2020).
2. N. Picqué and T. W. Hänsch, "Frequency comb spectroscopy," Nat. Photonics **13**, 146–157 (2019).
3. T. Udem, R. Holzwarth, and T. W. Hänsch, "Optical frequency metrology," Nature **416**, 233–237 (2002).
4. T. M. Fortier, M. S. Kirchner, F. Quinlan, *et al.*, "Generation of ultrastable microwaves via optical frequency division," Nat. Photonics **5**, 425–429 (2011).
5. A. L. Gaeta, M. Lipson, and T. J. Kippenberg, "Photonic-chip-based frequency combs," Nat. Photonics **13**, 158–169 (2019).
6. L. Chang, S. Liu, and J. E. Bowers, "Integrated optical frequency comb technologies," Nat. Photonics **16**, 95–108 (2022).
7. Y. Okawachi, B. Y. Kim, M. Lipson, *et al.*, "Chip-scale frequency combs for data communications in computing systems," Optica **10**, 977–995 (2023).
8. S. Biasi, G. Donati, A. Lugnan, *et al.*, "Photonic neural networks based on integrated silicon microresonators," Intell. Comput. **3**, 0067 (2024).
9. K. Han, D. A. Long, S. M. Bresler, *et al.*, "Low-power, agile electro-optic frequency comb spectrometer for integrated sensors," Optica **11**, 392–398 (2024).
10. H. H. Yoon, H. A. Fernandez, F. Nigmatulin, *et al.*, "Miniaturized spectrometers with a tunable van der Waals junction," Science **378**, 296–299 (2022).
11. B. Corcoran, M. Tan, X. Xu, *et al.*, "Ultra-dense optical data transmission over standard fibre with a single chip source," Nat. Commun. **11**, 2568 (2020).
12. Y.-S. Jang, S. Eom, J. Park, *et al.*, "Programmable spectral shaping for nanometric precision of frequency comb mode-resolved spectral interferometric ranging," Opt. Laser Technol. **170**, 110324 (2024).
13. Y. Ma, X. Huang, X. Wang, *et al.*, "Precise pulse shaping for quantum control of strong optical transitions," Opt. Express **28**, 17171–17187 (2020).
14. D. Lee, T. Nakamura, A. J. Metcalf, *et al.*, "Sub-GHz resolution line-by-line pulse shaper for driving superconducting circuits," APL Photonics **8**, 086115 (2023).
15. H.-H. Lu, M. Liscidini, A. L. Gaeta, *et al.*, "Frequency-bin photonic quantum information," Optica **10**, 1655–1671 (2023).
16. G. Moille, X. Lu, J. Stone, *et al.*, "Fourier synthesis dispersion engineering of photonic crystal microrings for broadband frequency combs," Commun. Phys. **6**, 144 (2023).
17. M. Roy, Z. Xiao, C. Dong, *et al.*, "Fundamental bandwidth limits and shaping of frequency-modulated combs," Optica **11**, 1094–1102 (2024).
18. H. Shu, L. Chang, Y. Tao, *et al.*, "Microcomb-driven silicon photonic systems," Nature **605**, 457–463 (2022).
19. L. M. Cohen, K. Wu, K. V. Myilswamy, *et al.*, "Silicon photonic microresonator-based high-resolution line-by-line pulse shaping," Nat. Commun. **15**, 7878 (2024).
20. T. Ozawa, H. M. Price, N. Goldman, *et al.*, "Synthetic dimensions in integrated photonics: From optical isolation to four-dimensional quantum Hall physics," Phys. Rev. A **93**, 043827 (2016).
21. D. Yu, W. Song, L. Wang, *et al.*, "Comprehensive review on developments of synthetic dimensions," Photonics Insights **4**, R06 (2025).
22. N. Englebert, N. Goldman, M. Erkintalo, *et al.*, "Bloch oscillations of coherently driven dissipative solitons in a synthetic dimension," Nat. Phys. **19**, 1014–1021 (2023).
23. A. Dikopoltsev, I. Heckelmann, M. Bertrand, *et al.*, "Collective quench dynamics of active photonic lattices in synthetic dimensions," Nat. Phys. **21**, 1134–1140 (2025).
24. E. Lustig, S. Weimann, Y. Plotnik, *et al.*, "Photonic topological insulator in synthetic dimensions," Nature **567**, 356–360 (2019).
25. L. Yuan, Y. Shi, and S. Fan, "Photonic gauge potential in a system with a synthetic frequency dimension," Opt. Lett. **41**, 741 (2016).

26. D. Cheng, K. Wang, C. Roques-Carmes, *et al.*, “Non-Abelian lattice gauge fields in photonic synthetic frequency dimensions,” Nature **637**, 52–56 (2025).
27. T. Ozawa and I. Carusotto, “Anomalous and quantum Hall effects in lossy photonic lattices,” Phys. Rev. Lett. **112**, 133902 (2014).
28. L. J. Maczewsky, K. Wang, A. A. Dovgiy, *et al.*, “Synthesizing multi-dimensional excitation dynamics and localization transition in one-dimensional lattices,” Nat. Photonics **14**, 76–81 (2020).
29. E. Lustig, L. J. Maczewsky, J. Beck, *et al.*, “Photonic topological insulator induced by a dislocation in three dimensions,” Nature **609**, 931–935 (2022).
30. Y. Hu, C. Reimer, A. Shams-Ansari, *et al.*, “Realization of high-dimensional frequency crystals in electro-optic microcombs,” Optica **7**, 1189–1194 (2020).
31. A. Schwartz and B. Fischer, “Laser mode hyper-combs,” Opt. Express **21**, 6196–6204 (2013).
32. U. A. Javid, R. Lopez-Rios, J. Ling, *et al.*, “Chip-scale simulations in a quantum-correlated synthetic space,” Nat. Photonics **17**, 883–890 (2023).
33. Z. Yang, E. Lustig, G. Harari, *et al.*, “Mode-locked topological insulator laser utilizing synthetic dimensions,” Phys. Rev. X **10**, 011059 (2020).
34. A. Tusnin, A. Tikan, K. Komagata, *et al.*, “Nonlinear dynamics and Kerr frequency comb formation in lattices of coupled microresonators,” Commun. Phys. **6**, 317 (2023).
35. A. Parriaux, K. Hammani, and G. Millot, “Electro-optic frequency combs,” Adv. Opt. Photonics **12**, 223–287 (2020).
36. J. B. Khurgin, “Energy and power requirements for alteration of the refractive index,” Laser Photonics Rev. **18**, 2300836 (2024).
37. M. Yu, D. Barton, III, R. Cheng, *et al.*, “Integrated femtosecond pulse generator on thin-film lithium niobate,” Nature **612**, 252–258 (2022).
38. M. Zhang, B. Buscaino, C. Wang, *et al.*, “Broadband electro-optic frequency comb generation in a lithium niobate microring resonator,” Nature **568**, 373–377 (2019).
39. A. K. Tusnin, A. M. Tikan, and T. J. Kippenberg, “Nonlinear states and dynamics in a synthetic frequency dimension,” Phys. Rev. A **102**, 023518 (2020).
40. A. Hugi, G. Villares, S. Blaser, *et al.*, “Mid-infrared frequency comb based on a quantum cascade laser,” Nature **492**, 229–233 (2012).
41. L. A. Sterczewski, C. Frez, S. Forouhar, *et al.*, “Frequency-modulated diode laser frequency combs at 2 $\mu$m wavelength,” APL Photonics **5**, 076111 (2020).
42. B. Dong, M. Dumont, O. Terra, *et al.*, “Broadband quantum-dot frequency-modulated comb laser,” Light Sci. Appl. **12**, 182 (2023).
43. I. Heckelmann, M. Bertrand, A. Dikopoltsev, *et al.*, “Quantum walk comb in a fast gain laser,” Science **382**, 434–438 (2023).
44. B. Marzban, L. Miller, A. Dikopoltsev, *et al.*, “A quantum walk comb source at telecommunication wavelengths,” arXiv (2024).
45. B. Meng, M. Singleton, J. Hillbrand, *et al.*, “Dissipative Kerr solitons in semiconductor ring lasers,” Nat. Photonics **16**, 142–147 (2021).
46. N. Opačak, D. Kazakov, L. L. Columbo, *et al.*, “Nozaki-Bekki solitons in semiconductor lasers,” Nature **625**, 685–690 (2024).
47. N. Opačak, F. Pilat, D. Kazakov, *et al.*, “Spectrally resolved linewidth enhancement factor of a semiconductor frequency comb,” Optica **8**, 1227–1230 (2021).
48. A. Dutt, M. Minkov, Q. Lin, *et al.*, “Experimental band structure spectroscopy along a synthetic dimension,” Nat. Commun. **10**, 3122 (2019).
49. N. Opačak, S. D. Cin, J. Hillbrand, *et al.*, “Frequency comb generation by bloch gain induced giant Kerr nonlinearity,” Phys. Rev. Lett. **127**, 093902 (2021).
50. H. A. Haus, “A theory of forced mode locking,” IEEE J. Quantum Electron. **11**, 323–330 (1975).
51. T. Zeng, Y. Dikmelik, F. Xie, *et al.*, “Ultrabroadband air-dielectric double-chirped mirrors for laser frequency combs,” Light Sci. Appl. **14**, 280 (2025).
52. U. Senica, A. Forrer, T. Olariu, *et al.*, “Planarized THz quantum cascade lasers for broadband coherent photonics,” Light Sci. Appl. **11**, 347 (2022).
53. Z. Wang, K. Van Gasse, V. Moskalenko, *et al.*, “A III–V-on-Si ultra-dense comb laser,” Light Sci. Appl. **6**, e16260 (2017).
54. I. Carusotto and C. Ciuti, “Quantum fluids of light,” Rev. Mod. Phys. **85**, 299–366 (2013).
55. I. Heckelmann, D. Pinto, U. Schmitt, *et al.*, “Fast and compact time-resolved spectroscopy enabled by Quantum Walk Combs,” arXiv (2025).
56. P. Täschler, M. Bertrand, B. Schneider, *et al.*, “Femtosecond pulses from a mid-infrared quantum cascade laser,” Nat. Photonics **15**, 919–924 (2021).
57. D. Cheng, E. Lustig, K. Wang, *et al.*, “Multi-dimensional band structure spectroscopy in the synthetic frequency dimension,” Light Sci. Appl. **12**, 158 (2023).
58. D. Piciocchi and A. Dikopoltsev, “Frequency comb shaping through staggered phase flux in fast gain lasers,” arXiv (2025).